\documentclass[12pt]{emulateapj}
\usepackage{natbib,xcolor}
\usepackage[caption=false]{subfig}    
\usepackage[normalem]{ulem}

\newcommand{\eg}{e.g.,}

\newcommand{\msun}{$M_\odot$}

\newcommand{\wise}{{\it WISE}}
\newcommand{\spitzer}{{\it Spitzer}}

\newcommand{\Mtwo}{M_{\mbox{\scriptsize 200}}}

\newcommand{\Nclword}{five}
\newcommand{\madcows}{MaDCoWS}
\newcommand{\scarma}{$\sigma_{\small \rm CARMA}$}
\newcommand{\swise}{$\sigma_{\small \rm WISE}$}
\newcommand{\sirsz}{$\sigma_{\small \rm IR-SZ}$}
\newcommand{\cirsz}{$\Delta C_{\small \rm IR-SZ}$}

\def\spose#1{\hbox to 0pt{#1\hss}}
\def\simlt{\mathrel{\spose{\lower 3pt\hbox{$\mathchar"218$}}
     \raise 2.0pt\hbox{$\mathchar"13C$}}}
\def\simgt{\mathrel{\spose{\lower 3pt\hbox{$\mathchar"218$}}
     \raise 2.0pt\hbox{$\mathchar"13E$}}}

 \shorttitle{\madcows. III. SZ Masses of Clusters at $z\sim 1$}
 \shortauthors{Brodwin et al.}

\begin{document}


\title{The Massive and Distant Clusters of \wise\ Survey.  III.
  Sunyaev-Zel'dovich Masses of Galaxy Clusters at $z\sim1$}

\author{M.~Brodwin\altaffilmark{1},
 C.~H.~Greer\altaffilmark{2},
 E.~M.~Leitch\altaffilmark{3,4},
 S.~A.~Stanford\altaffilmark{5,6},
 A.~H.~Gonzalez\altaffilmark{7},
 D.~P.~Gettings\altaffilmark{7},
 Z.~Abdulla\altaffilmark{3,4},
 J.~E.~Carlstrom\altaffilmark{3,4},
 B.~Decker\altaffilmark{1},
 P.~R.~Eisenhardt\altaffilmark{8},
 H.~W.~Lin\altaffilmark{9},
 A.~B.~Mantz\altaffilmark{3,4},
 D.~P.~Marrone\altaffilmark{2},
 M.~McDonald\altaffilmark{10},
 B.~Stalder\altaffilmark{9},
 D.~Stern\altaffilmark{8} \&
 D.~Wylezalek\altaffilmark{11}
}


\altaffiltext{1}{Department of Physics and Astronomy, University of Missouri, Kansas City, MO 64110}
\altaffiltext{2}{Steward Observatory, University of Arizona, Tucson, AZ 85121}
\altaffiltext{3}{Department of Astronomy and Astrophysics, University of Chicago, Chicago, IL 60637}
\altaffiltext{4}{Kavli Institute for Cosmological Physics, University of Chicago, IL 60637}
\altaffiltext{5}{University of California, Davis, CA 95616}
\altaffiltext{6}{Lawrence Livermore National Laboratory, Livermore, CA 94551}
\altaffiltext{7}{Department of Astronomy, University of Florida, Gainesville, FL 32611}
\altaffiltext{8}{Jet Propulsion Laboratory, California Institute of Technology, Pasadena, CA 91109}
\altaffiltext{9}{Harvard--Smithsonian Center for Astrophysics, Cambridge, MA 02138}
\altaffiltext{10}{Kavli Institute for Astrophysics and Space Research, Massachusetts Institute of Technology, Cambridge, MA 02139}
\altaffiltext{11}{European Southern Observatory, Garching bei M\"unchen, Germany}


\begin{abstract}

  We present CARMA 30 GHz Sunyaev-Zel'dovich (SZ) observations of
  \Nclword\ high-redshift ($z\ga 1$), infrared-selected galaxy
  clusters discovered as part of the all-sky Massive and Distant
  Clusters of \wise\ Survey (\madcows).  The SZ decrements measured
  toward these clusters demonstrate that the MaDCoWS selection is
  discovering evolved, massive galaxy clusters with hot intracluster
  gas.  Using the SZ scaling relation calibrated with South Pole
  Telescope clusters at similar masses and redshifts, we find these
  \madcows\ clusters have masses in the range $\Mtwo \approx 2-6
  \times10^{14} $ \msun.  Three of these are among the most massive
  clusters found to date at $z\ga 1$, demonstrating that \madcows\ is
  sensitive to the most massive clusters to at least $z = 1.3$.  The
  added depth of the AllWISE data release will allow all-sky infrared
  cluster detection to $z\approx 1.5$ and beyond.

\end{abstract}





\section{Introduction}

The last decade, roughly since the launch of the {\it Spitzer Space
  Telescope}, has been a remarkably productive time for the discovery
of high-redshift ($z \ga 1$) galaxy clusters.  This is due in large
part to \spitzer's superb sensitivity to massive galaxies out to very
high redshift, but also to the maturation of Sunyaev-Zel'dovich (SZ)
effect surveys and increasingly sophisticated X-ray and radio active
galactic nucleus (AGN)-based surveys.  These techniques have
successfully identified galaxy clusters beyond $z>1$
\citep[\eg][]{rosati04, mullis05, stanford06, eisenhardt08, muzzin09,
  brodwin10, galametz10, brodwin13, zeimann13, reichardt13,
  hasselfield13, bleem15} and even at $z \ga 1.5$
\cite[\eg][]{fassbender11_sfr, santos11, stanford12, zeimann12,
  muzzin13, bayliss14, willis13, newman14}.

Despite their successes in discovering high-redshift clusters, deep
X-ray \citep[\eg][]{fassbender11, mehrtens12} and \spitzer\ surveys
\citep[\eg][]{eisenhardt08, papovich10, rettura14} are limited to
relatively small areas ($< 100$ deg$^2$), and thus do not probe the
volume required to meaningfully sample the high-mass end of the $z \ga
1$ cluster mass function.  The South Pole Telescope
\citep[SPT,][]{reichardt13, bleem15} and Atacama Cosmology Telescope
\citep[ACT,][]{hasselfield13} SZ surveys, though much larger, are
still limited to a few thousand square degrees, whereas the all-sky
{\it Planck} SZ survey \citep{planck13_SZ} is limited to $z < 1 $ due
to its large beam.

The Massive and Distant Clusters of \wise\ Survey
\citep[\madcows,][]{gettings12, stanford14} is a new IR-selected
galaxy cluster survey based on the all-sky catalogs of the {\it
  Wide-field Infrared Survey Explorer} \citep[\wise,][]{wright10}.
The combination of \wise\ infrared and Sloan Digital Sky Survey (SDSS)
DR8 optical photometry \citep{sdss_dr8} allows us to robustly isolate
galaxy clusters at $z \ga 1$ in the northern hemisphere. The first
spectroscopically confirmed \madcows\ cluster, MOO\ J2342+1301 at
$z=0.99$, was reported by \citet{gettings12}.  The reader is referred
to that paper, along with the upcoming survey paper (Gonzalez et
al.~in prep), for a more complete description of the survey
methodology.

The current 10,000 deg$^2$ survey footprint is four times larger than
the SPT-SZ survey \citep{bleem15} and 1000 times larger than the area
of the IRAC Distant Cluster Survey (IDCS), in which the most massive
$z>1.5$ galaxy cluster known to date was found \citep{brodwin12,
  stanford12, gonzalez12}.  Given the unprecedented volume surveyed
{\it at high redshift}, the \madcows\ sample should contain a large
number of very massive, distant clusters.

In this paper we present Combined Array for Research in
Millimeter-wave Astronomy\footnote{http://www.mmarray.org} (CARMA) 30
GHz observations of \Nclword\ $z \ga 1$ \madcows\ clusters spanning a
range of infrared richnesses.  The spectroscopic confirmations for all
but one of these clusters are given in \citet{stanford14}.  In
\textsection{\ref{Sec: Data}} we present the \madcows\ clusters,
including a (red sequence) photometric redshift measurement for
MOO\ J1014+0038, and describe the CARMA SZ observations.  In
\textsection{\ref{Sec: Masses}} we describe the measurements of total
Comptonization, from which we infer masses.  We discuss our results in
\textsection{\ref{Sec: Discussion}.  We assume a concordance
  $\Lambda$CDM cosmology with $\Omega_M = 0.3$, $\Omega_{\Lambda} =
  0.7$ and $H_0 = 70$ km s$^{-1}$ Mpc$^{-1}$.  \spitzer/IRAC and
  ground-based optical magnitudes are calibrated to the Vega and AB
  systems, respectively.

\section{Data}
\label{Sec: Data}
\subsection{CARMA Observations}

CARMA is an interferometer that consists of six 10.4~m, nine 6.1~m,
and eight 3.5~m telescopes, providing fields of view of FWHM 3.8\arcmin,
6.6\arcmin\ and 11.4\arcmin\ at 30 GHz, respectively.  All 23 telescopes
have 30 and 90 GHz receivers, while the 10.4~m and 6.1~m telescopes
have an additional 230 GHz receiver.  CARMA is equipped with two
correlators: an 8-station correlator with 7.5 GHz of bandwidth per
baseline (the ``wideband correlator'') and a more flexible correlator
that can be configured to correlate 23 stations with 2 GHz bandwidth
per baseline at 30~GHz (``spectral line correlator'').  To maximize
sensitivity, both correlators can be used simultaneously.

Clusters MOO\ J0012+1602, MOO\ J0319-0025 and MOO\ J1014+0038 were
observed when the 10.4~m and 6.1~m telescopes were in E configuration,
the 3.5~m telescopes were in SL configuration, and the signals were
processed by both correlators (``E+SL''). MOO\ J1155+3901 was observed
using the wideband correlator and the eight 3.5 m telescopes in the SH
configuration. MOO\ J1514+1346 was observed twice, first using the
eight 3.5~m telescopes in the SL configuration with the wideband
correlator and later with all 23 telescopes in the E+SL configuration
using only the spectral line correlator.  All observations were
centered around 31 GHz.

With the exception of MOO\ J1514+1346, cluster observations in the
E+SL configuration used the wideband correlator to process the
intermediate frequency (IF) signal from eight of the 6.1~m telescopes.
The synthesized beam formed by the 6.1~m $\times$ 6.1~m baselines is
approximately 1\arcmin, a resolution well matched to the angular size
of the SZ signal from these distant clusters. The wideband correlator
provides most of the sensitivity to the cluster signal in these
observations. The 23-element observations of MOO\ J1514+1346 lacked
the wideband correlator due to a hardware problem and were primarily
useful for confirming a point source that was not well detected by the
SL data alone. For the standard E+SL observations, the spectral line
correlations sample baselines from 0.35-12.0~k$\lambda$, while the wideband
correlations sample baselines from 0.65-4.0~k$\lambda$.

\begin{figure*}[bthp]
    \centering
    \subfloat{\plotone{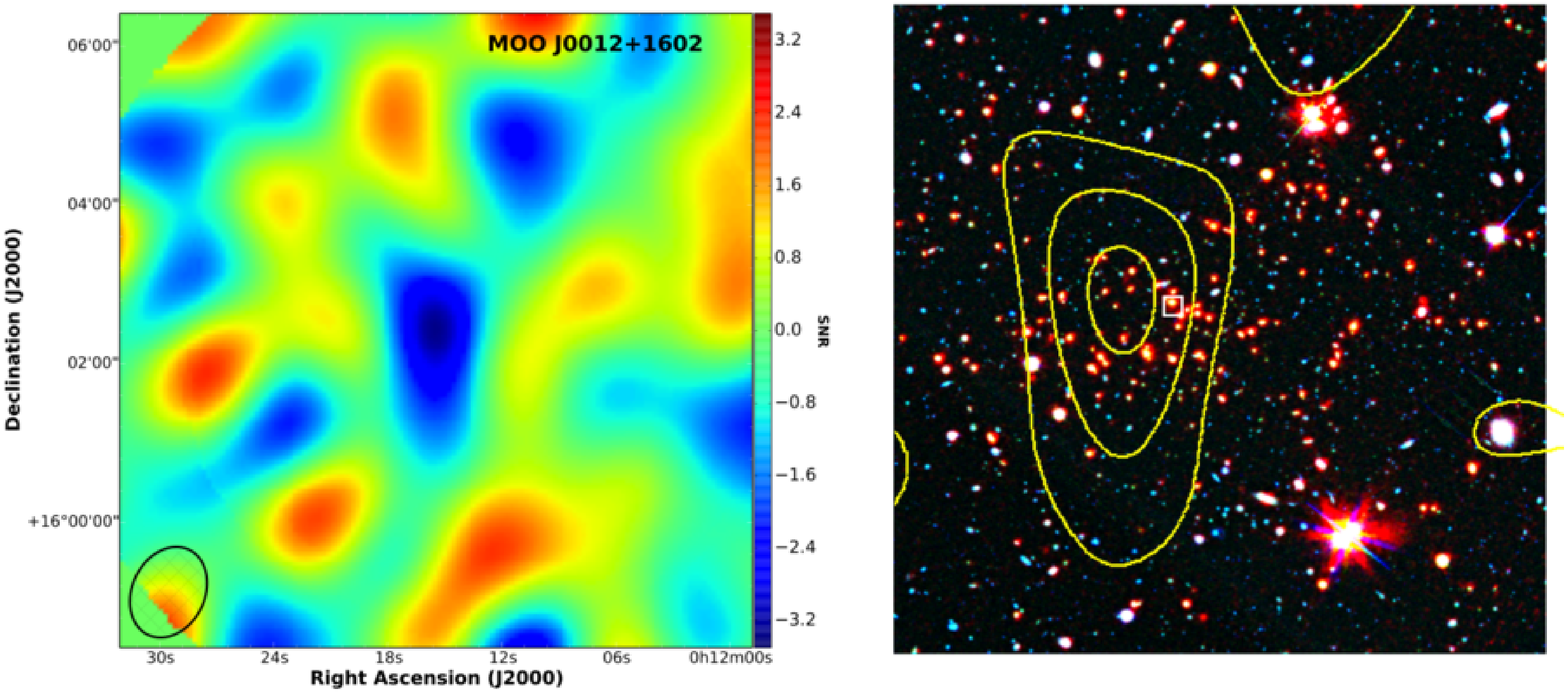}}\\
    \subfloat{\plotone{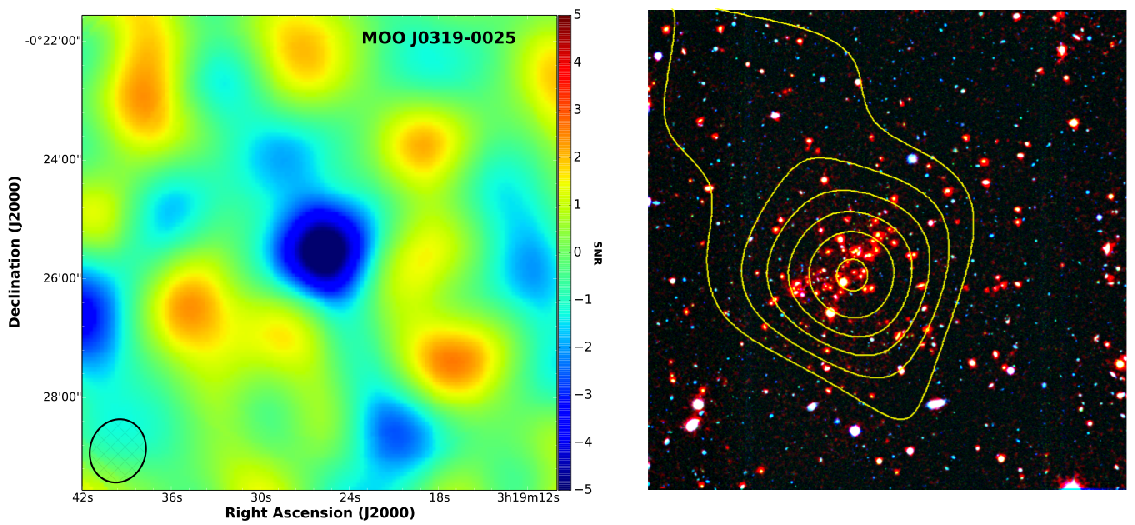}}\\
    \subfloat{\plotone{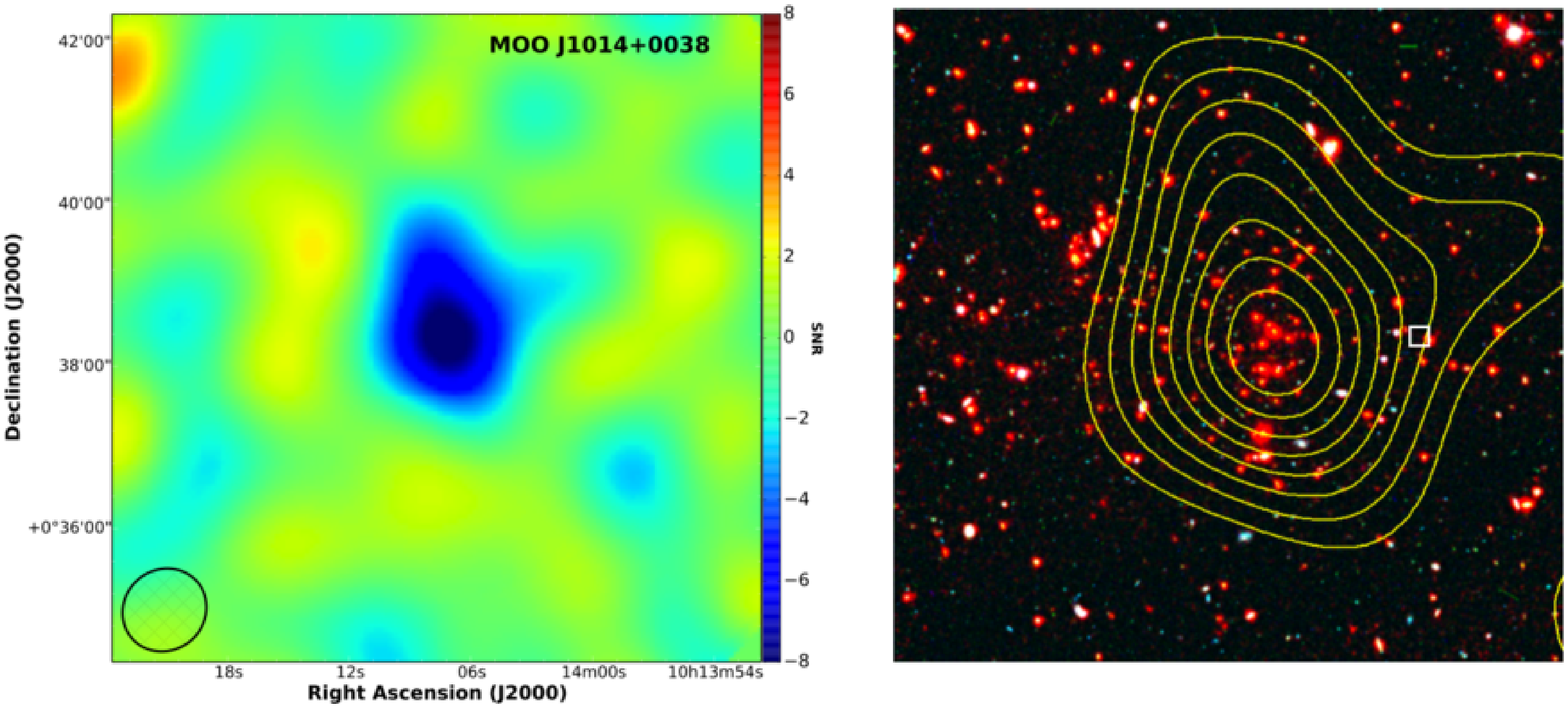}}\\
    \caption{$8\arcmin \times 8\arcmin$ CARMA 30 GHz maps ({\it left
        column}) and $4\arcmin \times 4\arcmin$ optical/IR images
      ({\it right column}) of high-redshift \madcows\ clusters.  The
      CARMA maps are in units of SNR in the SZ detection, which are
      negative to reflect the decrement.  The FWHM of the synthesized
      beams are shown in the lower left in each map.  The optical
      layers of the pseudo-color images are composed of $r$ and $z$
      images from Gemini/GMOS-N, with the exception of MOO\ J1014+0038
      for which we have these same bands from Magellan/IMACS.  The IR
      layer is IRAC 3.6 $\mu$m except for MOO\ J1155+3901, for which
      we use the \wise\ W1 band at 3.4 $\mu$m.  Contours of the SZ
      decrements in SNR are overplotted on the optical/IR images.  In
      all cases the least significant contour is SNR = $-1$ and the
      contours increase in significance by $\Delta$SNR = $-1$.  The
      locations of the emissive sources list in Table \ref{Tab: agn}
      are indicated with white boxes.}
\label{Fig: images}
\end{figure*}

\begin{figure*}[bthp]
    \centering
    \ContinuedFloat
    \subfloat{\plotone{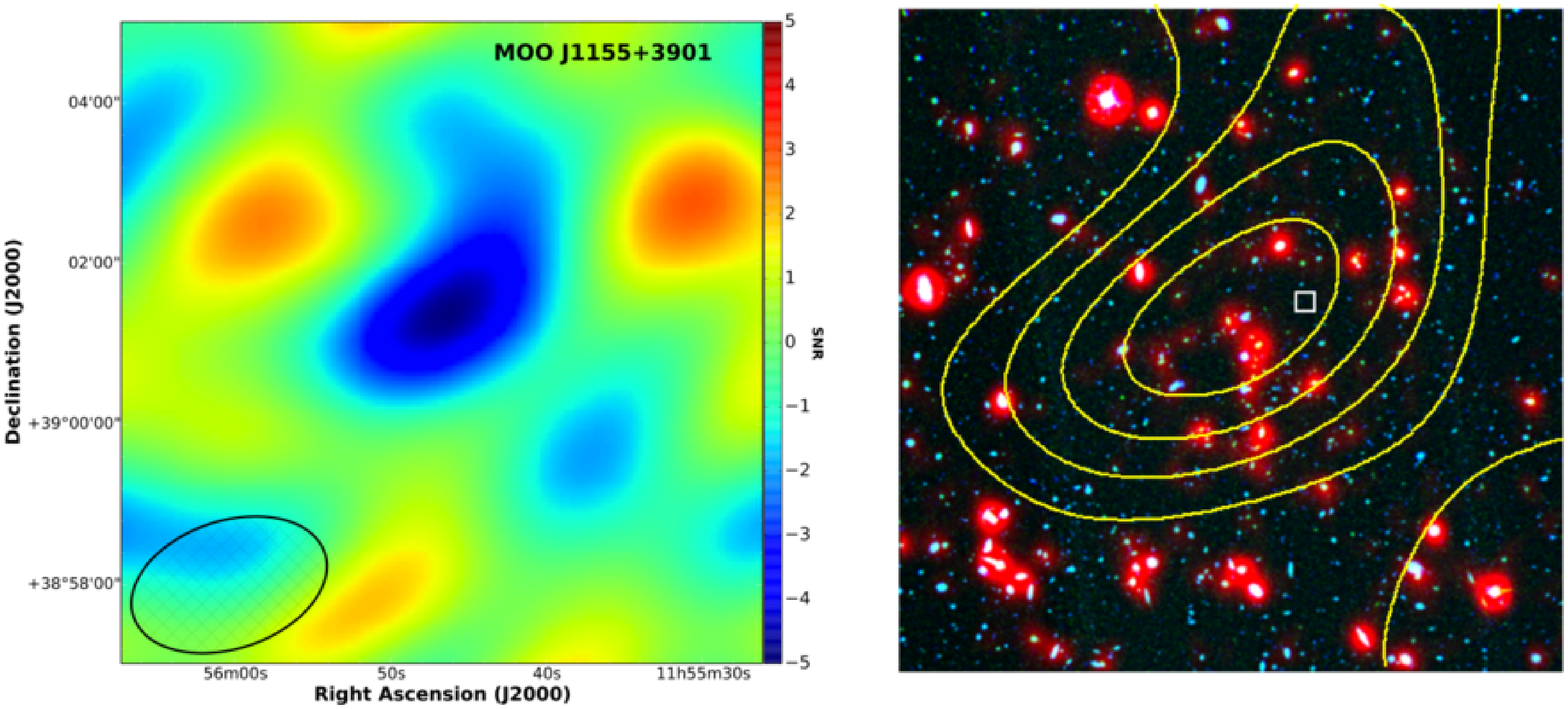}}\\
    \subfloat{\plotone{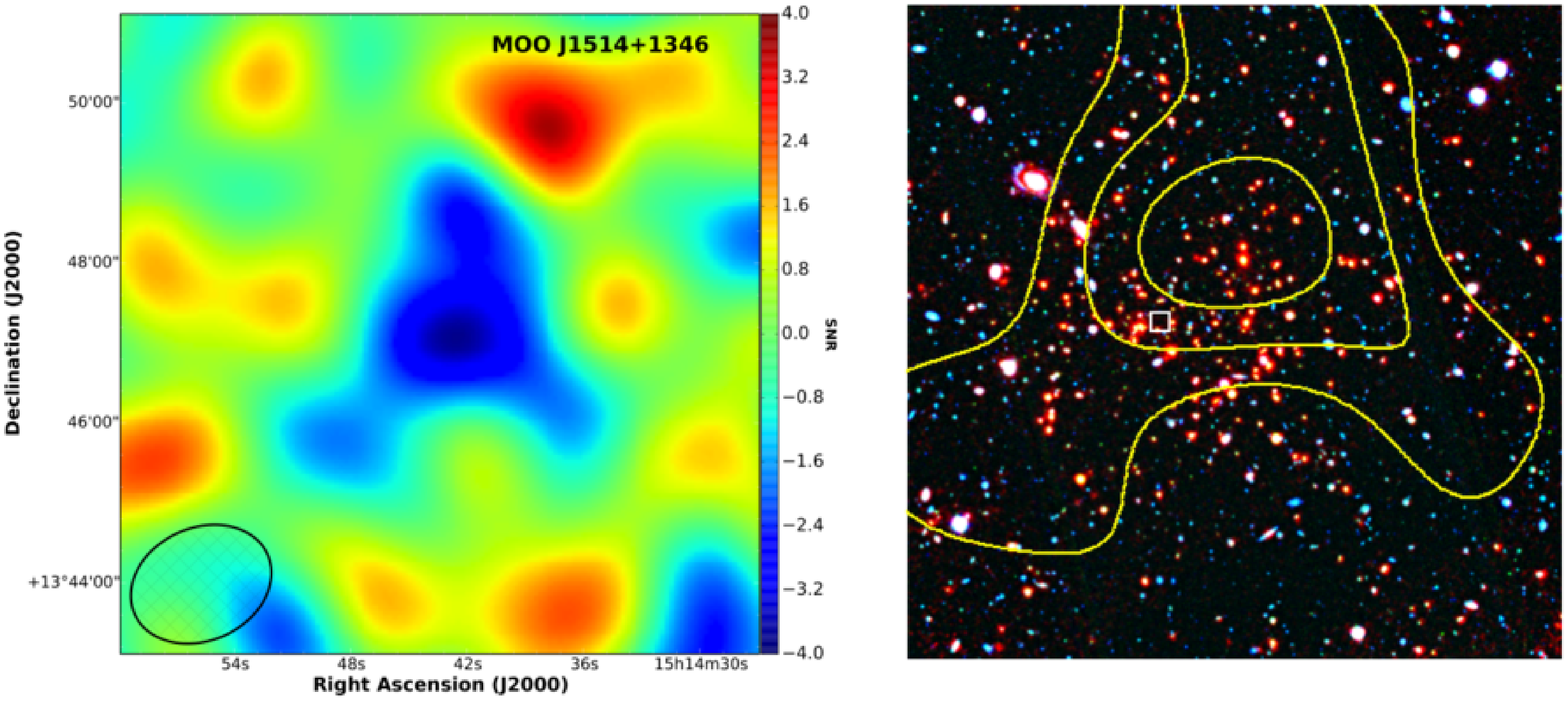}}\\
    \caption{Continued.}
\end{figure*}

The CARMA observations, summarized in Table~\ref{Tab: sample}, used
the WISE centroid positions as the pointing and phase centers. The
data were reduced with a pipeline using MIRIAD \citep{sault95} similar
in function to the one described in \citet{muchovej07}.  After
filtering for bad weather and instrumentation problems, the data were
gain-calibrated using observations of bright, unresolved quasars
interleaved every 15 minutes between the cluster observations.  Flux
densities are calibrated against observations of Mars using the model
presented in \citet{rudy87}. The pipeline produces a set of
flux-calibrated visibilities of the cluster field.  The CARMA images
in Figure \ref{Fig: images} are created by combining
naturally-weighted $uv$ data from both correlators.  They are in SNR
units and a 2 k$\lambda$ cutoff is applied to the data to highlight
the cluster-sensitive baselines in the images.  Models fit to emissive
sources are removed (see \textsection\ref{Sec: sources}). The residual
maps are CLEANed inside a square box 3\arcmin\ on a side, centered on
the position of the map with the largest absolute value in SNR
units. CLEAN is allowed to proceed until the largest peak in the map
is 1.5 times the map RMS value.

\begin{deluxetable*}{lcccccc}
  \tabletypesize{\normalsize} \tablecaption{CARMA Observations of
    \madcows\ Clusters \label{Tab: sample}} \tablewidth{0pt} \tablehead{
    \colhead{Cluster ID} & \colhead{R.A.} & \colhead{Decl.} &
    \colhead{UT Dates} & \colhead{CARMA Array} &     \colhead{Exp.~Time\tablenotemark{1}} &
    \colhead{Map RMS Noise\tablenotemark{2}} \\
    \colhead{} & \colhead{(J2000)} & \colhead{(J2000)} &
    \colhead{} &\colhead{}& \colhead{(hr)} & \colhead{(mJy)} } \startdata
  MOO\ J0012$+$1602 &  00:12:13.0 &   $+$16:02:15 &   2013 Sep 24; Oct 1,3,6  &   E+SL &  6.0 & 0.11   \\
  MOO\ J0319$-$0025 &  03:19:24.4 &   $-$00:25:21 &   2013 Sep 30             &   E+SL &  1.0 & 0.26   \\
  MOO\ J1014$+$0038 &  10:14:08.4 &   $+$00:38:26 &   2013 Oct 6-7            &   E+SL &  2.2 & 0.17   \\
  MOO\ J1155$+$3901 &  11:55:45.6 &   $+$39:01:15 &   2012 May 11-12          &   SH   &  7.2 & 0.33   \\
  MOO\ J1514$+$1346 &  15:14:42.7 &   $+$13:46:31 &   2013 Jun 1,3,5-7,9,11   &   SL   &  8.4 & 0.20   \\ 
                    &             &               &   2013 Aug 15             &   E+SL &  0.5 & 0.19   
\enddata
\tablenotetext{1}{Total on-source exposure time, excluding overhead,
  calibrations and flagged (unused) data.}  \tablenotetext{2}{The map
  RMS noise is the mean noise within a 3.5\arcmin\ radius circle
  centered on the pointing center.}
\end{deluxetable*}

\begin{deluxetable}{lccc}
  \tabletypesize{\normalsize} \tablecaption{Emissive
    Sources \label{Tab: agn}} \tablewidth{0pt} \tablehead{
    \colhead{Cluster ID} &   \colhead{R.A.} & \colhead{Decl.}  & \colhead{Flux}  \\
    \colhead{} & \colhead{(J2000)} & \colhead{(J2000)} &
    \colhead{(mJy)}} \startdata
  MOO\ J0012$+$1602 & 00:12:14.39 &   $+$16:02:23.6     &    $1.91 \pm 0.07$  \\
  MOO\ J0319$-$0025 &      -      &         -           &    -  \\
  MOO\ J1014$+$0038 & 10:14:03.84 &   $+$00:38:26.0     &    $0.31 \pm 0.07$  \\
  MOO\ J1155$+$3901 & 11:55:44.36 &   $+$39:01:28.6     &    $0.72 \pm 0.33$  \\
  MOO\ J1514$+$1346 & 15:14:33.59 &   $+$13:52:47.3     &    $\,\,8.04 \pm 0.93$\tablenotemark{$\dagger$} \\
                     & 15:14:44.57 &   $+$13:46:34.8     &    $0.18 \pm 0.07$
\enddata
\tablenotetext{$\dagger$}{This is an extended radio source 6\arcmin\ NW of the
  phase center.  The flux was measured in an elliptical aperture with a
  16\arcsec\ semi-major axis and an axial ratio of 0.72.}
\end{deluxetable}

\subsection{Gemini Data}

All but one of these clusters (MOO\ J1014+0038) were observed with
the Gemini Multi-Object Spectrograph (GMOS) on Gemini-North.  Exposure
times of 15 min were obtained in the $r$- and $z$-bands to produce
color magnitude diagrams (CMDs) from which red-sequence members could
be selected for follow-up spectroscopic observations.  These images
are combined with IRAC 3.6 $\mu$m or \wise\ 3.4 $\mu$m images to make
the color images for four of the \Nclword\ clusters shown in Figure
\ref{Fig: images}.

These images were used to construct several of the masks, for both
Gemini/GMOS-N and Keck/DEIMOS, with which we spectroscopically
confirmed 20 MaDCoWS clusters to date, including four of the \Nclword\
in the present work.  \citet{stanford14} provide a full description of
the MaDCoWS spectroscopy.

\begin{figure}[bthp]
\epsscale{1.15}
\plotone{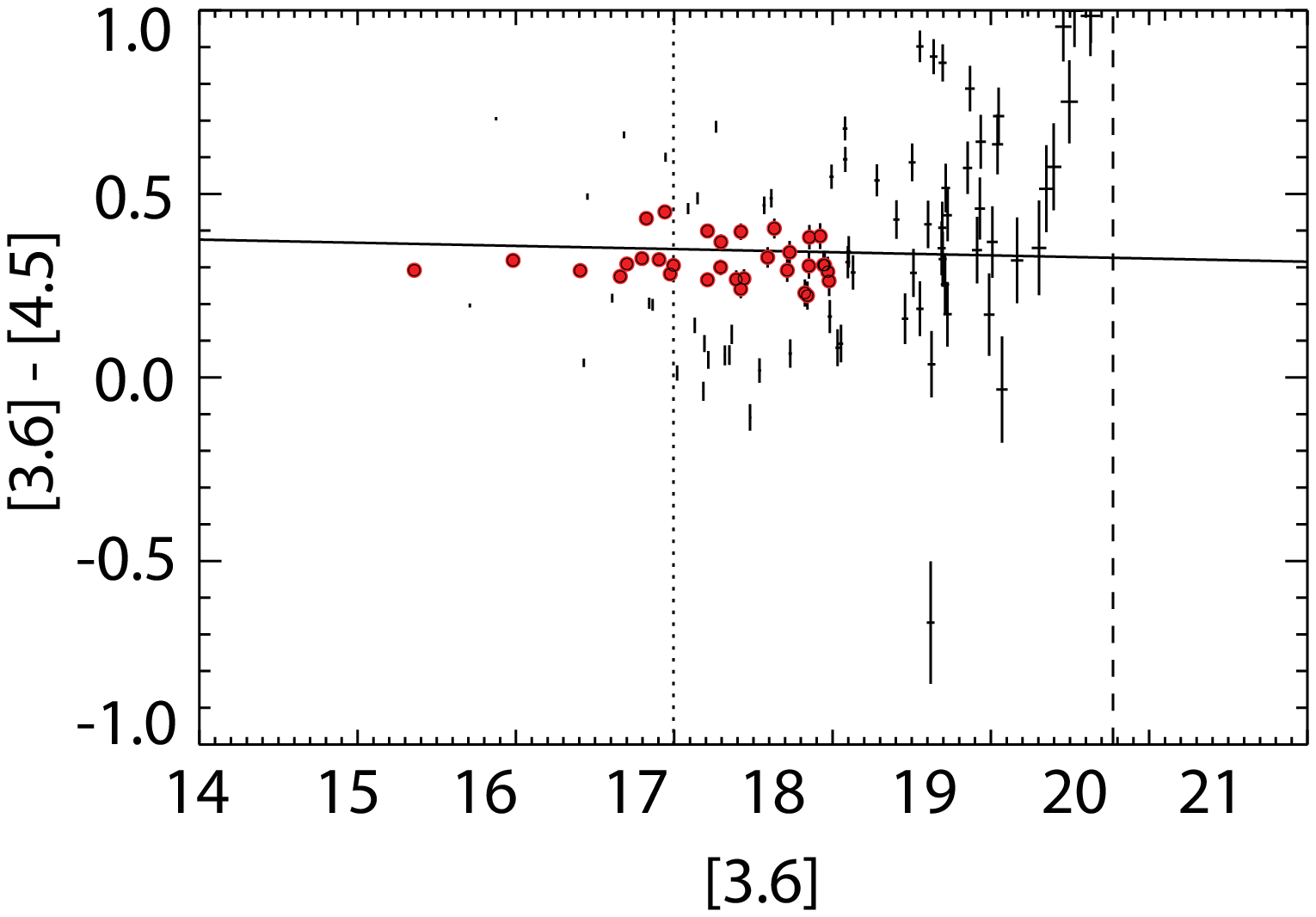}
\vspace*{0.15cm}
\plotone{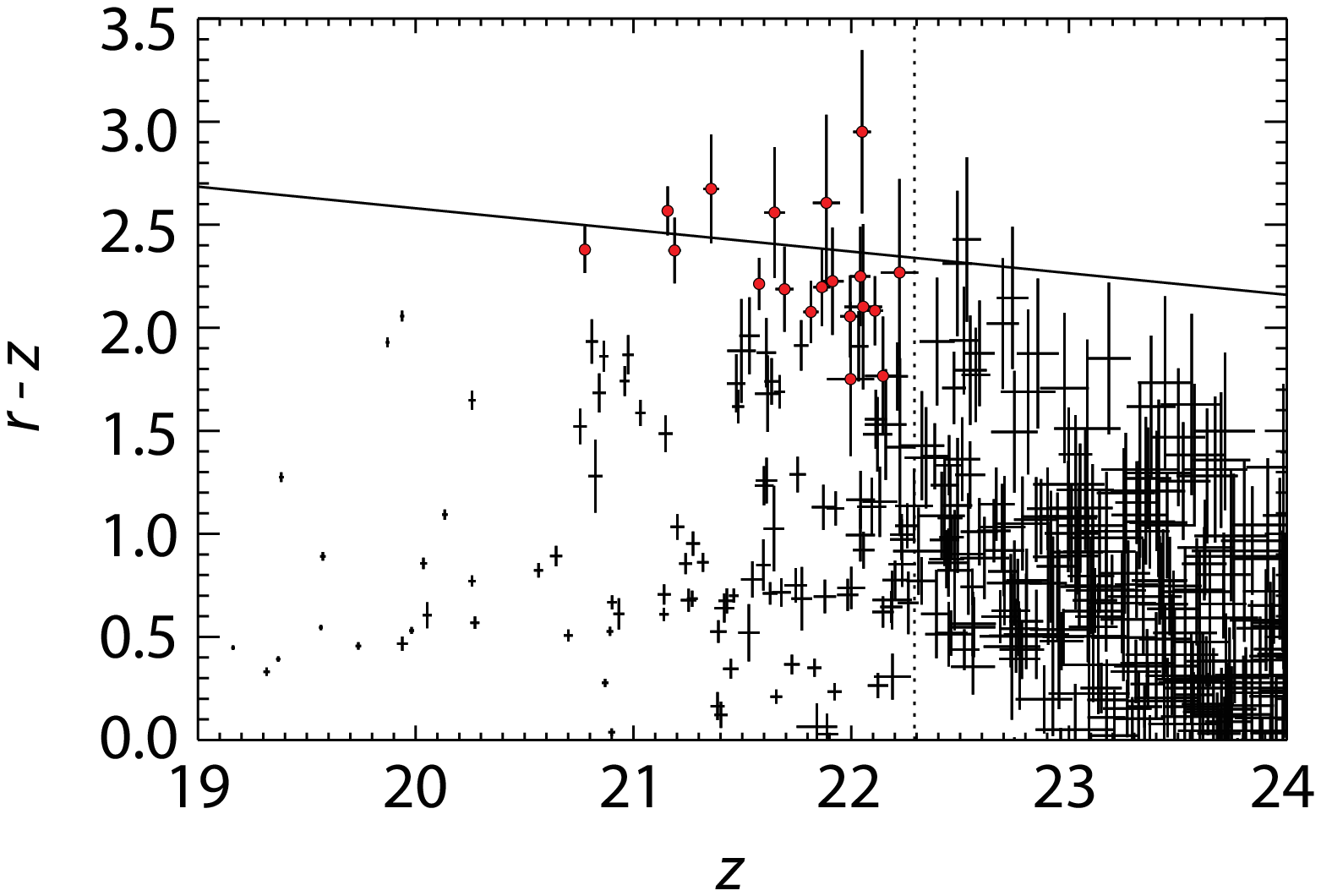}
\caption{IRAC (upper panel) and optical (lower panel) color-magnitude
  diagrams (CMDs) for cluster MOO\ J1014+0038, based on IRAC and
  Magellan/IMACS data, respectively.  The red points show the galaxies
  that have colors consistent with the best-fit red sequence
  model. The vertical dotted lines are the $M*$ magnitudes in the 3.6
  $\mu$m and $z$ bands, respectively, at the best-fit photometric
  redshift. The dashed vertical line in the top panel is the detection
  limit in the IRAC data.  Using the method described in
  \citet{song12}, we find a best-fit photometric redshift of $1.27 \pm
  0.08$ is consistent with both CMDs.}
\label{Fig: cmd}
\end{figure}

\subsection{Magellan Data}

Cluster MOO\ J1014+0038 was observed on UT 2014 January 22 with the
Inamori Magellan Areal Camera and Spectrograph
\citep[IMACS;][]{dressler06} in the $g$, $r$ and $z$ bands for 2, 12,
and 12 minutes, respectively.  These data were reduced with the SPT
optical pipeline, as described in \citet{song12}, and the $r$ and $z$
images are combined with IRAC 3.6 $\mu$m to make the color image of
this cluster shown in Figure \ref{Fig: images}.

Photometric redshifts based on the cluster red-sequence were measured
from both optical and IRAC color magnitude relations, shown in Figure
\ref{Fig: cmd}.  The best-fit photometric redshift in both cases is
consistent with $z_{\rm ph} = 1.27 \pm 0.08$.  We have used this value
in converting the observed SZ decrement to a total mass.

\section{Analysis}
\label{Sec: Masses}

\subsection{Identification of Compact Radio Sources}
\label{Sec: sources}

Compact radio sources are a potential source of uncertainty in SZ
cluster mass measurements.  Since these synchrotron sources are often
variable, it is important to identify and remove their contribution
contemporaneously with the SZ measurement.  The long baselines of our
CARMA observations (up to 12 k$\lambda$) can identify compact emission
sources at high sensitivity.  We identify contaminating sources by
visual inspection of the long-baseline data and checking the 1.4~GHz
catalogs from the NRAO VLA Sky Survey \citep[NVSS,][]{condon98} and
the VLA Faint Images of the Radio Sky at Twenty-Centimeters
\citep[FIRST,][]{becker95}.  A radio source is jointly modeled with
the cluster signal if its flux is detected in the CARMA data at 3
$\sigma$ or greater, or if there is a source in the NVSS or FIRST
catalog. In the case that the source is only marginally-detected in
the CARMA data and present in a VLA catalog, its position is fixed and
only the flux is allowed to vary.  The simultaneous fit mitigates the
contamination of our significance and mass fits due to variable
emissive sources, as described below.

In MOO\ J1155$+$3901, which has a centrally located source of
emission detected at $\sim2.5\sigma$ in the CARMA data, there exists a
slight positive correlation between the measured $Y$ parameter for the
cluster and the point source flux. In the remaining 4 clusters, there
is no sign of covariance between the cluster parameters and other
source parameters.  The coordinates and beam-corrected fluxes of
compact radio sources identified toward each cluster are given in
Table \ref{Tab: agn}.

\subsection{SZ Decrement Significance}

To determine the significances of the observed SZ decrements, we
measure the difference in $\chi^2$ between a model of our observations
that contains no cluster with models (described in
\textsection\ref{Sec: masses}}) that do include a cluster. In the
cases where there is contaminating radio emission, models for these
point sources are fitted in both the cluster and no-cluster
models. The models are fit to the data in the $uv$-plane using a
Markov chain Monte Carlo routine, which correctly accounts for the
noise in our data from the heterogeneous CARMA array \citep[see,
e.g.,][]{plagge13}.

The resulting $\chi^2$ values are converted to significances in terms
of Gaussian standard deviations.  They range from 2.7 $\sigma$ to 9.5
$\sigma$ and are listed in Table \ref{Tab: masses}.  In fitting the SZ
cluster centroids, a uniform 1.5\arcmin\ radius prior centered on the
IR position is used.  For low significance decrements, negative noise
spikes can formally bias the SNR to higher values.  However, given the
strong prior of an IR-selected, confirmed cluster at the targeted
position, and the small number of independent beams ($<10$) over which
the SZ centroid is allowed to vary, we expect all of these decrements
to be robust.  Using an approximate analytical calculation of the
noise properties in our CARMA maps we conservatively estimate the
probability of false detection for the two least significant clusters
to be $\sim$8-10\%. The higher significance clusters have no
significant probability of false detection.

\subsection{SZ Mass Measurements}
\label{Sec: masses}
The mass estimates are produced following the method described in
recent CARMA papers \citep[e.g.][]{brodwin12}. Briefly, we
parameterize the SZ signal as a pressure profile that we integrate to
measure the integrated Compton $Y$ parameter.  We use a generalized
NFW pressure profile as presented in \citet{nagai07},
\begin{equation}
P(x) = \frac{P_0}{x^\gamma (1+x^\alpha)^{(\beta-\gamma)/\alpha}},
\end{equation}
where $P_0$ is the normalization, $x\equiv r/r_s$ is a dimensionless
radial variable, and $P_0$ and $r_s$ are allowed to vary.  The
power-law exponents are fixed to the ``universal'' values
(i.e.,~$\alpha = 1.0510$, $\beta = 5.4905$, and $\gamma = 0.3091$) of
\citet{arnaud10}. The cluster centroid and the positions and fluxes of
any coincident emissive sources are allowed to vary as well. There is
a strong degeneracy between $P_0$ and $r_s$ in the data, however the
resulting $Y$ parameter, as defined below, is well-constrained.

We integrate the derived pressure profile to a cutoff radius to
calculate the integrated $Y$ parameter,
\begin{equation}
Y_\Delta = \frac{1}{D_A^2}\frac{\sigma_T}{m_e c^2} \int_0^{r_{\Delta}} P(r/r_s) dV.
\end{equation}
To determine the cutoff radius, we enforce consistency with the
$Y_{SZ,500}-M_{500}$ scaling relation derived in \citet{andersson11}
by requiring that the chosen integration radius (and thus, mass) and
resulting $Y$ lie on the mean relation. We determine the final
integration radius iteratively, and the value of $Y$ typically
converges in roughly five iterations.  The $M_{500}$ values in
Table~\ref{Tab: masses} correspond to the derived values of $r_{500}$
in our cosmology.

\begin{deluxetable*}{llcrrccc}
  \tabletypesize{\normalsize} \tablecaption{SZ Properties of \madcows\
    Clusters \label{Tab: masses}} \tablewidth{0pt} \tablehead{
    \colhead{ID} & \colhead{Redshift} & \colhead{Significance} &
    \colhead{$r_{500}$} &\colhead{$Y_{500}$} &\colhead{$M_{500}$} &
    \colhead{$M_{200}$\tablenotemark{1}} &   \colhead{\cirsz \tablenotemark{2}} \\
    \colhead{} & \colhead{} & \colhead{($\sigma$)} &
    \colhead{(Mpc)} &\colhead{($10^{-6}$ Mpc$^{2}$)} & \colhead{($10^{14}$ \msun)} &
    \colhead{($10^{14}$ \msun)}& \colhead{(Mpc)}} \startdata 
  MOO\ J0012$+$1602 &    0.944                 & 2.7 &  $0.56 \pm 0.07$ &  $7.6  \pm 4.6$ &  $1.4 \pm 0.5$    &  $2.2 \pm 0.8$ &  0.270    \\
  MOO\ J0319$-$0025 &    1.194                 & 6.6 &  $0.65 \pm 0.03$ &  $30.0 \pm 5.9$ &  $3.1 \pm 0.4$    &  $5.1 \pm 0.6$ &  0.194    \\
  MOO\ J1014$+$0038 &    $1.27 \pm 0.08$\tablenotemark{3}       & 9.5 &  $0.66 \pm 0.02$ &  $37.0 \pm 7.1$ &  $3.4 \pm 0.4$    &  $5.6 \pm 0.6$ &  0.142    \\
  MOO\ J1155$+$3901 &    1.009                 & 2.8 &  $0.69 \pm 0.06$ &  $26 \pm 11$    &  $2.9 \pm 0.7$    &  $4.7 \pm 1.2$ &  0.099    \\
  MOO\ J1514$+$1346 &    1.059                 & 3.2 &  $0.61 \pm 0.05$ &  $16.0 \pm 6.9$ &  $2.2 \pm 0.6$    &  $3.5 \pm 0.9$ &  0.282
\enddata
\tablenotetext{1}{$M_{200}$ masses were extrapolated from the measured
  $M_{500}$ masses using the \citet{duffy08} mass--concentration
  relation.}
\tablenotetext{2}{\cirsz\ is the offset between the IR and SZ positional centroids.}

\tablenotetext{3}{The fit for this cluster assumed a Gaussian redshift
  prior with $\sigma_z = 0.08$ centered on the photometric redshift of
  $z=1.27$.  Due to the flat angular diameter distance, the redshift
  independence of the SZ and the narrow evolutionary window, no
  scatter is imposed by the redshift uncertainty at the significance
  presented.}
\end{deluxetable*}
\color{black}

\subsection{SZ and IR Centroids}

The SZ and IR galaxy density centroids trace different physical probes
of the potential, namely the integrated pressure of the ICM and the
distribution of massive galaxies, respectively.  For individual
clusters these need not be coincident.  Indeed, recent work has
suggested that even X-ray and SZ centroids may be offset from each
other during a major merger \citep{zhang14}.

\begin{figure}[bthp]
\epsscale{1.15}
\plotone{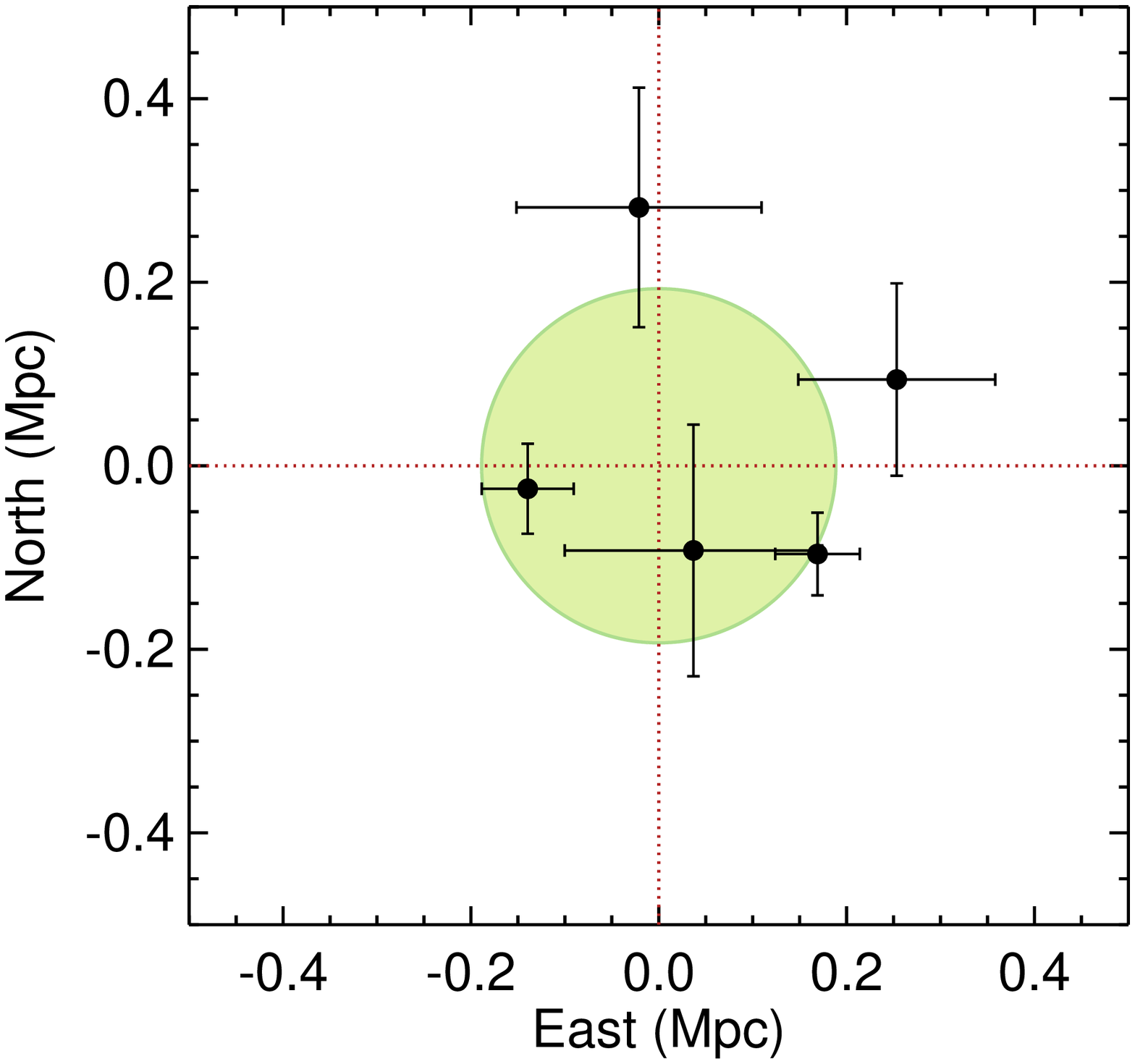}
\caption{Positional offsets between SZ and \wise\ IR cluster
  positions.  The errors bars include the positional uncertainties of
  the \wise\ and CARMA observations.  The average residual centroid
  difference, likely due to physical differences between the IR galaxy
  density and SZ ICM centroid measurements, is shown by the filled
  circle.}
\label{Fig: offset}
\end{figure}

In Figure \ref{Fig: offset} we plot the angular offsets of the CARMA
SZ cluster centroids from the \wise\ IR cluster positions, taken as
the peaks in the wavelet detection map.  The SZ-IR offset is less than
300 kpc for all the clusters in this sample.  The error bars represent
the quadrature sum of the uncertainties in the CARMA and \wise\
centroids.  The CARMA positional errors, \scarma, determined in the
fit described in \textsection{\ref{Sec: masses}}, are jointly fit with
the positions and fluxes of the emissive radio sources.  The \wise\
centroiding precision is limited by finite gridding (15\arcsec/pixel)
in the \wise\ cluster search, and to a lesser extent, to confusion
caused by the relatively large \wise\ beam ($\sim 6\arcsec$).
However, the metric offsets caused by these technical factors should
be small (\swise\ $\la 50$ kpc) for these high-SNR cluster detections.

The larger offsets seen here are likely due to real differences
between IR galaxy density and SZ ICM centroids.  We estimate the
average residual positional offset, \sirsz, by setting the reduced
chi-squared statistic equal to unity:
\begin{equation}
  \chi^2_\nu = \frac{1}{\nu} \sum_{\rm clusters}\left(\frac{\rm
      Offset}{\sqrt{\sigma_{\small \rm CARMA}^2 + \sigma_{\small \rm
          \it WISE}^2 + \sigma_{\small \rm IR-SZ}^2}}\right)^2 = 1,
\end{equation}
where $\nu$ is the number of degrees of freedom.  We find \sirsz\ $ =
188.6$ kpc, shown as the filled circular region at the center of the
plot.  A similar result is obtained using IR centers defined by the
BCGs identified in follow-up IRAC imaging, confirming the IR-SZ offset
is not due to an unknown systematic in the \wise\ centering.

\section{Discussion}
\label{Sec: Discussion}

We have presented SZ decrements and derived masses for \Nclword\
distant ($z\ga 1$), massive ($\Mtwo \approx 2 - 6 \times 10^{14}$
\msun) MaDCoWS galaxy clusters selected via their stellar mass
signatures in the \wise\ All-Sky data release. Four of these are
spectroscopically confirmed \citep{stanford14}, and we have presented
a reliable photometric redshift for the fifth derived from both deep
optical and \spitzer\ photometry.

The current MaDCoWS catalog is drawn from the 10,000 deg$^2$ overlap
between the \wise\ and SDSS surveys.  With the largest volume yet
surveyed at high redshift, we expect to find very rare, massive
clusters in the distant Universe.  Indeed, three of the clusters in
this preliminary CARMA pilot study have masses in excess of $\Mtwo >
4.5 \times 10^{14}$ \msun, among the most massive discovered to date
at $z\ga 1$.  More extensive characterization of the MaDCoWS sample is
underway with CARMA and other facilities, including an ongoing AO-13
{\it XMM-Newton} program targeting several of the clusters presented in
this work.

\begin{figure*}[bthp]
\epsscale{1.15}
\plotone{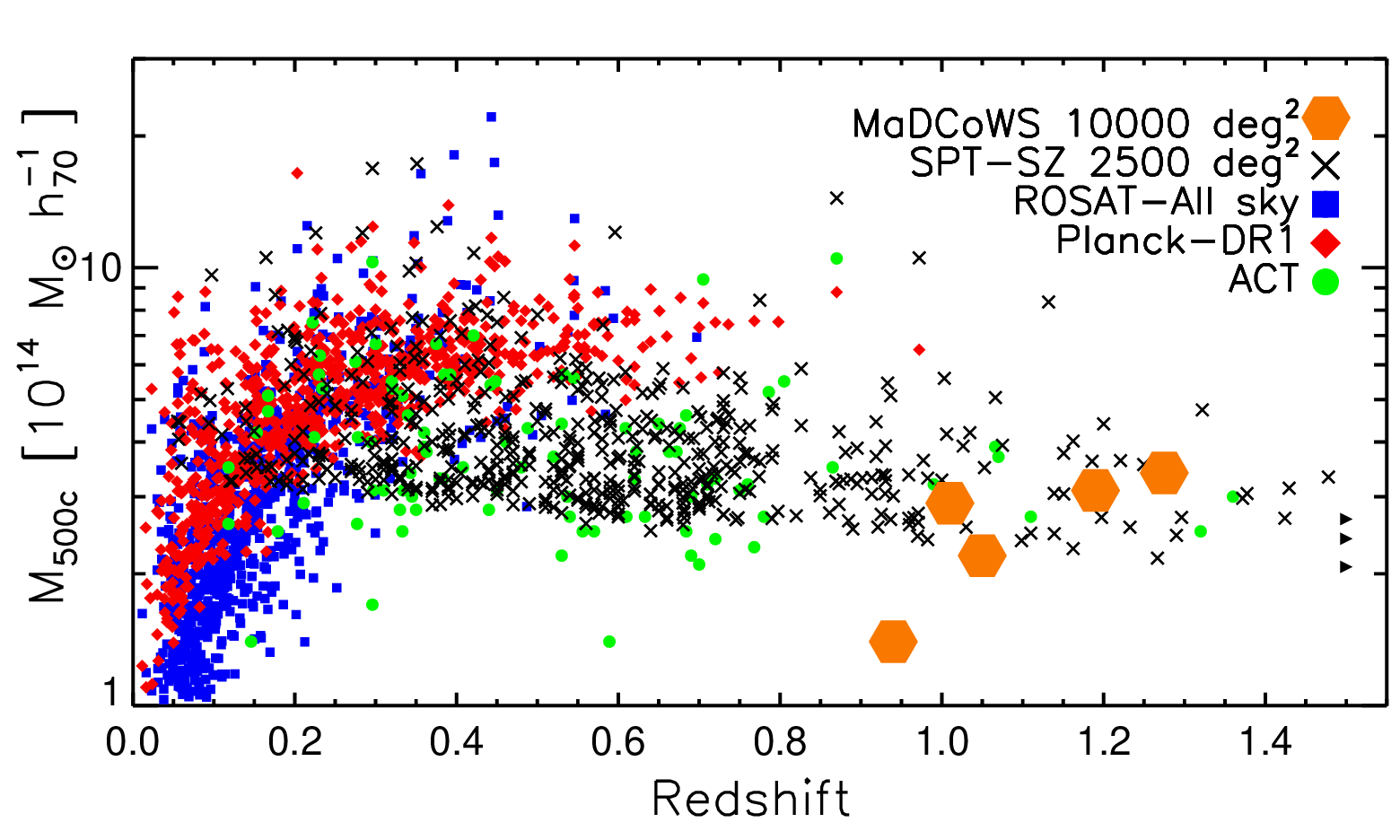}
\caption{Comparison of the mass-redshift plane for MaDCoWS clusters
  with those of other wide-angle cluster surveys, including {\it
    ROSAT} \citep{Piffaretti11}, {\it Planck} \citep{planck13_SZ}, ACT
  \citep{marriage11, hasselfield13} and SPT \citep{reichardt13,
    bleem15}.  The SPT masses were computed using a slightly different
  scaling relation than the \citet{andersson11} relation adopted here;
  to more directly compare the MaDCoWS masses with the SPT sample the
  former should be increased by $\sim 8\%$ (L.~Bleem, priv. comm.).
  The MaDCoWS clusters are comparable in mass to the high-redshift
  SZ-selected clusters from SPT and ACT. This figure is adapted from
  \citet{bleem15}.}
\label{Fig: mass_z}
\vspace*{0.25cm}
\end{figure*}

Figure \ref{Fig: mass_z}, adapted from \citet{bleem15}, shows the
mass--redshift plane for the largest, wide-area cluster surveys to
date.  These include the {\it ROSAT} X-ray surveys
\citep{Piffaretti11}, composed of the NORAS \citep{bohringer00},
REFLEX \citep{bohringer04} and MACS
\citep{ebeling01,ebeling07,ebeling10} cluster catalogs, and the SZ
cluster catalogs from the {\it Planck} \citep{planck13_SZ}, ACT
\citep{marriage11, hasselfield13} and SPT \citep{reichardt13, bleem15}
collaborations.  The IR-selected MaDCoWS clusters presented in this
work, shown as orange hexagons, are similar to the high-redshift
clusters selected from the high-resolution SZ surveys (i.e., SPT and
ACT).  They are drawn from the massive cluster population at $0.9 < z
< 1.3$, and as ICM and/or weak lensing masses for the larger MaDCoWS
sample are measured, we should identify high-redshift ($z \ge 1$)
clusters even more massive than those seen to date in the smaller-area
SZ surveys.

In the near future, the MaDCoWS sample will be extended in both area
and redshift. The recent AllWISE data release \citep{cutri13} has
significantly more uniform sky coverage, with an average of twice the
exposure time over the previous all-sky catalog.  This should allow
the identification of massive galaxy clusters to $z \approx 1.5$.  Our
current search is also limited to the SDSS footprint.  With the
arrival of large optical surveys in, or extending to, the Southern
hemisphere, such as Pan-STARRS, VST and DES, our search area will soon
cover the bulk of the extragalactic sky.  This will enable the
discovery, within MaDCoWS, of the rarest, most massive clusters at $z
\ga 1$.

\acknowledgments We thank the anonymous referee for helpful comments
that improved the paper.  We thank L.~Bleem for providing the code and
data to produce Figure \ref{Fig: mass_z} and B.~Benson for helpful
conversations.  Support for CARMA construction was derived from the
Gordon and Betty Moore Foundation, the Kenneth T. and Eileen L. Norris
Foundation, the James S. McDonnell Foundation, the Associates of the
California Institute of Technology, the University of Chicago, the
states of California, Illinois, and Maryland, and the National Science
Foundation. Ongoing CARMA development and operations are supported by
the National Science Foundation under a cooperative agreement, and by
the CARMA partner universities; the work at Chicago was supported by
NSF grant AST- 1140019. Additional support was provided by PHY-
0114422.  This publication makes use of data products from the
Wide-field Infrared Survey Explorer, which is a joint project of the
University of California, Los Angeles, and the Jet Propulsion
Laboratory/California Institute of Technology, funded by the National
Aeronautics and Space Administration.  M.B., D.P.G., A.H.G.\ and
S.A.S.\ acknowledge support for this research from the NASA
Astrophysics Data Analysis Program (ADAP) through grant NNX12AE15G.
This work was supported by a NASA Keck PI Data Award, administered by
the NASA Exoplanet Science Institute.  This work is based in part on
observations made with the {\it Spitzer Space Telescope}, which is
operated by the Jet Propulsion Laboratory, California Institute of
Technology under a contract with NASA.  This work is based in part on
data obtained at the W.~M.~Keck Observatory, which is operated as a
scientific partnership among the California Institute of Technology,
the University of California and the National Aeronautics and Space
Administration.  The Observatory was made possible by the generous
financial support of the W.~M.~Keck Foundation.  Based in part on
observations obtained at the Gemini Observatory, which is operated by
the Association of Universities for Research in Astronomy, Inc., under
a cooperative agreement with the NSF on behalf of the Gemini
partnership: the National Science Foundation (United States), the
National Research Council (Canada), CONICYT (Chile), the Australian
Research Council (Australia), Minist\'{e}rio da Ci\^{e}ncia,
Tecnologia e Inova\c{c}\~{a}o (Brazil) and Ministerio de Ciencia,
Tecnolog\'{i}a e Innovaci\'{o}n Productiva (Argentina).  Data were
obtained in Program IDs GN-2013A-Q-44 and GN-2013B-Q-8.  This paper
includes data gathered with the 6.5 meter Magellan Telescopes located
at Las Campanas Observatory, Chile.

\newpage
\bibliographystyle{astron2} \bibliography{bibfile}

\end{document}